# A tiling algorithm for the aperiodic monotile Tile(1,1)

Henning U. Voss

Cornell MRI Facility, Cornell University, Ithaca, NY
Department of Radiology, Weill Cornell Medicine, New York, NY

June 7, 2024. Edited November 1, 2024.

*Abstract* — An algorithm is provided to tile the plane with the aperiodic monotile Tile(1,1) recently discovered by Smith et al. (2023). Their geometric composition guidelines are expanded into a numerical MATLAB algorithm. The intention is to remove a possible obstacle for researchers interested in applications of this fundamental tiling of the plane.

Recently, Smith et al. [1] discovered a family of aperiodic monotiles of the plane, T(a,b), with geometric parameters a and b. In particular, the special case of Tile(1,1) stands out, as it is a weakly chiral monotile; in other words, Tile(1,1) can tile the plane only in a nonperiodic fashion, and reflected (turned-over) tiles are not needed nor allowed. A relative of Tile(1,1), the so-called "Hat", is a non-chiral monotile and requires the inclusion of reflected tiles for the nonperiodic tiling of the plane [2]. A slight generalization of Tile(1,1), the "Specter", consists of a family of Tile(1,1) monotiles with deformed edges. The Specter is a strictly chiral aperiodic monotile, and reflected tiles are allowed (but do not contribute to the tiling).

Smith et al. [1] provide geometric composition guidelines that can be used to assemble the tiling in an iterative way. A graphical web app is provided by C.S. Kaplan at https://cs.uwaterloo.ca/~csk/spectre/app.html, also allowing for downloading the coordinates. A published MATLAB algorithm to automatically generate a tiling with Tile(1,1) seems to be lacking. The provided algorithm solves this problem. It can construct tilings with Tile(1,1) for an, in principle, arbitrary number of tiles, limited only by computing power. Visual verification of the correctness of the algorithm was performed up to the $8^{th}$ iteration of tilings, consisting of 16,908,641 tiles.

The algorithm translates the geometric composition guidelines of the tiling into a numeric algorithm. Details of the implementation including some non-trivial procedures that might not be evident from the geometric guidelines can be best understood by consulting the annotated MATLAB script provided in the Appendix. It is described briefly as follows.

The goal is to tile the plane with the aperiodic monotile Tile(1,1). The geometric guidelines achieve this by using "Specter" clusters S and "Mystic" clusters M. S and M consist of one and two tiles Tile(1,1), respectively, to seed the iteration. From one iteration to the next, S and M are mirrored. Refer to Figure A1.2a in Smith et al. (2023). The final tiling of the plane is then given by the S cluster at the final iteration, to be set by the user. In the MATLAB script, first, two functions are defined, which are used to place tiles or clusters of tiles at specified locations and at specified angles, and to mirror the tiling from one iteration to the next. Then, the expected number of tiles for a tiling with the final number of iterations is computed, to provide an estimate of resources needed. The first pattern S0 = Tile(1,1) is defined, for initiating the iteration, and labelled with four key points. Then the Mystic pattern M0 is defined as a two-cluster of S0s. For the first two iterations, the key points used to compose the tiling of the next iteration are provided explicitly. They were read off from the beforementioned figure. The first rotation angles for placement of the initial tiles to generate S1 and S2 were obtained by trial and error. After composition of S1, its key points are cleared and four new key points are being defined to compose S2 from S1 in a similar way as S1 has been composed from S0. The rotation angles for S2 were obtained by trial and error. Luckily, for subsequent iterations they remained constant, up to iteration number 8. It is conjectured here that they converged at iteration 2 to their asymptotic value. This insight was not expected from the geometric guidelines. After S2 is complete, Sn (n > 2) are assembled by extracting the key points from the previous iteration.

The provided MATLAB algorithm "TileOneOne" is written in a self-explanatory way and commented. It should be very easy to convert it to any other programming langue. It should also be easy to use it with minor modifications to tile the plane with any of the Specters. The script is provided in the Appendix and can also be downloaded at https://github.com/henningle/TileOneOne.git.

Figure 1 shows the tiling for iteration 5, consisting of 34649 tiles. Figure 2 show the tiling for iteration 8, which has 16,908,641 tiles.

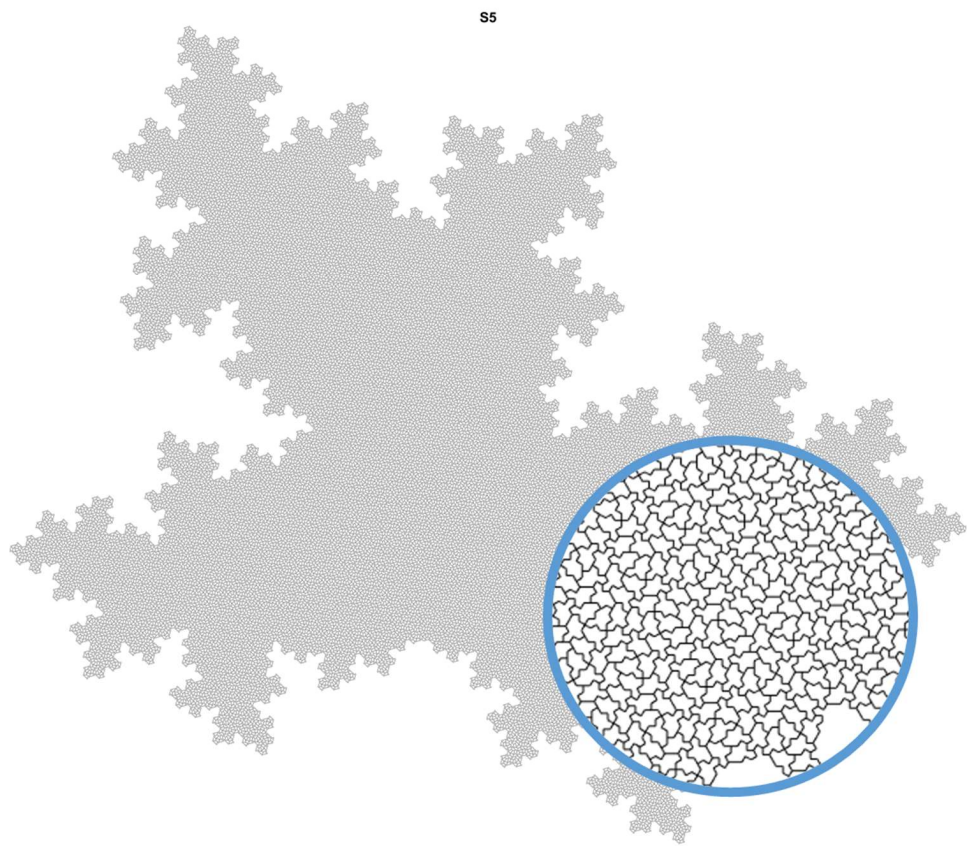

**Figure 1:** Tiling with the aperiodic monotile Tile(1,1) for the fifth iteration, shown with a "magnifying glass" to show the monotile components.

A tiling algorithm for Tile(1,1)

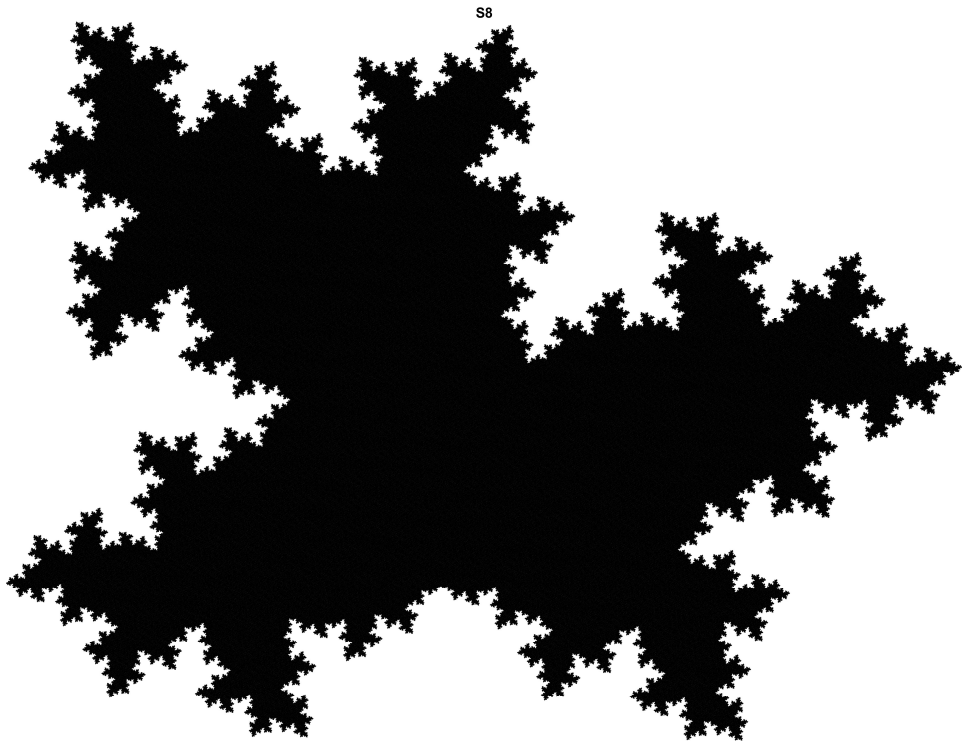

**Figure 2:** Tiling with the aperiodic monotile Tile(1,1) for the eighth iteration, highlighting the fractal boundary of the tiling set.

**Appendix**

The MATLAB algorithm TileOneOne.m.

```
%% TileOneOne.m, 6.6.2024
% To compose a tiling from the aperiodic monotile Tile(1,1)
% Please cite as
% "H.U. Voss, A tiling algorithm for the aperiodic monotile Tile(1,1), arXiv 2406.05236
(2024)"
% The license attached in GitHub applies, at https://github.com/henningle/TileOneOne.git

%% Parameters

maxN=4; % Maximum iteration. The largest maxN tested was maxN=8 (16,908,641 tiles)

plotkeypoints=true; % Set to false if plotting key points not wanted

%% Main script

% Tile placement function
% Each tile has 3 components: x, y, key point number (0 for no key point, 1 to 4 if key point)
% The placetile function takes in a matrix of size (number of rows, 3)
% The first 2 columns are the coordinates of the tiling, the third is a label of key points
placetile=@(x, angle, coord) ...
    [transpose( [cos(angle*pi/180), -sin(angle*pi/180); ...
    sin(angle*pi/180), cos(angle*pi/180)]*x(:,1:2)' ...
    +transpose(repmat(coord,[size(x,1),1]))) , x(:,3)];

mirror=@(x) [transpose([-1, 0; 0, 1]*(x(:,1:2)')),  x(:,3)];
```

A tiling algorithm for Tile(1,1)

```matlab
% Number of expected tiles for each iteration
nS=zeros(1,maxN+1);
nM=nS;
nS(1)=1; nM(1)=2; % This is index 0
for n=2:maxN+1
    nS(n)=nM(n-1)+7*nS(n-1);
    nM(n)=nM(n-1)+6*nS(n-1);
end
nS=nS(2:end);
nM=nM(2:end);
disp(['Predicted number of tiles for N = 1, 2, ...: ' num2str(nS)])

%% S0 and M0

% Some definitions
cos30=cos(30*pi/180); sin30=sin(30*pi/180);
cos60=cos(60*pi/180); sin60=sin(60*pi/180);
cos120=cos(120*pi/180); sin120=sin(120*pi/180);
cos150=cos(150*pi/180); sin150=sin(150*pi/180);
cos210=cos(210*pi/180); sin210=sin(210*pi/180);
cos240=cos(240*pi/180); sin240=sin(240*pi/180);
cos300=cos(300*pi/180); sin300=sin(300*pi/180);
cos330=cos(330*pi/180); sin330=sin(330*pi/180);

% One Tile(1,1)
% Each tile has 14 edges and 15 points, the last point for closure
% Plus one NaN for separation of tiles while plotting
% x-components of specter S
Stmp1=[ 0
    0
    cos30
    cos30+cos120
    cos30+cos120+cos60
    cos30+cos120+cos60+cos330
    cos30+cos120+cos60+cos330
    cos30+cos120+cos60+cos330+1
    cos30+cos120+cos60+cos330+1+cos300
    cos30+cos120+cos60+cos330+1+cos300+cos210
    cos30+cos120+cos60+cos330+1+cos300+cos210
    cos30+cos120+cos60+cos330+cos300+cos210
    cos30+cos120+cos60+cos330+cos300+cos210+cos240
    cos30+cos120+cos60+cos330+cos300+cos210+cos240+cos150
    cos30+cos120+cos60+cos330+cos300+cos210+cos240+cos150
    NaN];
% y-components of specter S
Stmp2=[0
    1
    1+sin30
    1+sin30+sin120
    1+sin30+sin120+sin60
    1+sin30+sin120+sin60+sin330
    sin30+sin120+sin60+sin330
    sin30+sin120+sin60+sin330
    sin30+sin120+sin60+sin330+sin300
    sin30+sin120+sin60+sin330+sin300+sin210
    sin30+sin120+sin60+sin330+sin300+sin210-1
    sin30+sin120+sin60+sin330+sin300+sin210-1
    sin30+sin120+sin60+sin330+sin300+sin210-1+sin240
```

A tiling algorithm for Tile(1,1)

```
        sin30+sin120+sin60+sin330+sin300+sin210-1+sin240+sin150
        sin30+sin120+sin60+sin330+sin300+sin210+sin240+sin150
        NaN];
% Specter S
S=[Stmp1,Stmp2];

% Augment S with key points to be carried over during iterative composition
keypoints=zeros(16,1); keypoints(4)=1; keypoints(6)=2; keypoints(8)=3; keypoints(14)=4;
S=[S,keypoints];
keyind=find(S(:,3)>0)'; % The indexes that point to key points
% S(keyind,3)' % Should be 1,2,3,4 for iteration 0

% The Mystic
M=[S; placetile(S,-30,[sqrt(3)/2,-1.5])];
M(:,3)=0; % M should not have key points not to contaminate S

%% Iteration loop

for N=1:maxN
    
    plotting=0; if N==maxN; plotting=1; end % Plot only the last iteration
    
    S=mirror(S);
    M=mirror(M);
    
    % The first two iterations require explicit key points for the next iteration
    switch N
        case 1
            Sind=[4,6,8,14]; % Key points on S0, for making S1
            Mind=Sind+16;
            angles=[150,90,90,30,-30,-30,-90];
        case 2
            Sind=[84,118,132,38]; % Key points on S1, for making S2
            % Indexing changes in M2 due to the missing patch
            Mind=[84-16,118-16,132-16,38];
            angles=[120,60,60,0,-60,-60,-120]; % Conjecture: Angles converged from here on
    end
    
    % For iteration 3 and higher, keypoints are computed from the previous two iterations
    if N > 2
        Sind=circshift(keyind,-1);
        Mind(1:3)=Sind(1:3)-16*nS(N-2);
    end
    
    % Define some tiling cells
    cell0=M;
    cell1=placetile([S(:,1:2)-S(Sind(3),1:2), S(:,3)],angles(1), M(Mind(3),1:2));
    cell2=placetile([S(:,1:2)-S(Sind(2),1:2), S(:,3)],angles(2),cell1(Sind(4),1:2));
    cell3=placetile([S(:,1:2)-S(Sind(3),1:2), S(:,3)],angles(3),cell2(Sind(1),1:2));
    cell4=placetile([S(:,1:2)-S(Sind(2),1:2), S(:,3)],angles(4),cell3(Sind(4),1:2));
    cell5=placetile([S(:,1:2)-S(Sind(2),1:2), S(:,3)],angles(5),cell4(Sind(4),1:2));
    cell6=placetile([S(:,1:2)-S(Sind(3),1:2), S(:,3)],angles(6),cell5(Sind(1),1:2));
    cell7=placetile([S(:,1:2)-S(Sind(2),1:2), S(:,3)],angles(7),cell6(Sind(4),1:2));
    
    %% Assemble S and M
    
    S=[cell0 % 2x16 elements because this is the mystic M
        cell1 % 16 elements
        cell2
```

A tiling algorithm for Tile(1,1)

```matlab
            cell3
            cell4
            cell5
            cell6
            cell7
            ];
    M=[cell0 % 2x16 elements
        cell1 % 16 elements
        cell2
        cell4 % Note we skipped cell3
        cell5
        cell6
        cell7
        ];
    M(:,3)=0; % Again, M should not have key points

    if plotting
        figure('position',[500.,100.,1000,1000]);
        plot(S(:,1),S(:,2),'k',LineWidth=.1);
        axis image
        title(['S' num2str(N)])
    end

    %% Key points of this iteration

    keyind=find(S(:,3)>0); % Always 7*4=28 elements after pruning previous iteration

    if plotting
        if plotkeypoints
            % Plot all keypoints before pruning, in red
            hold on
            for n=1:4:length(keyind)
                plot(S([keyind(n:n+3);keyind(n)],1),S([keyind(n:n+3);keyind(n)],2),'rO-');
            end
        end
    end

    %% Key points for next iteration

    % At the beginning, there are always 9*4=36 key points. At the end, there should be 4
    keyind2=keyind([2,13,22,25]); % Read off from Smith23b Fig. A1.2a and previous plot
    if N>1; keyind2=keyind([2,13,22,25]+1); end % Don't ask
    % S(keyind2,3)' % Should be (2,1,2,1) for iteration 1 and (3,2,3,2) for iteration >= 2

    S(:,3)=0; % Clean out this iteration's key points

    % Set new keypoints for composing the next iteration
    S(keyind2(1),3)=1;
    S(keyind2(2),3)=2;
    S(keyind2(3),3)=3;
    S(keyind2(4),3)=4;

    keyind=find(S(:,3)>0); % Must be 38 84 118 132 for iteration 2

    if plotting
        if plotkeypoints
            % Plot only key points taken over to next iteration, in blue
            plot(S([keyind;keyind(1)],1),S([keyind;keyind(1)],2),'bO-');
        end
```

A tiling algorithm for Tile(1,1)

```
    end

    % Count tiles
    nStrue=size(S,1)/16;
    disp(['Iteration ' num2str(N) ' has ' num2str(nStrue) ' tiles'])

end

%% Save tiling to file

S=S(:,1:2); % Saving only the lattice proper
save(['TileOneOne_Ni' num2str(N) '.mat'], "S");
```